# Quantitative and Qualitative Performance Evaluation of Commercial Metal Artifact Reduction Methods: Dosimetric Effects on the Treatment Planning

Mohammad Ghorbanzadeh[1], Seyed Abolfazl Hosseini[1], Bijan Vosoughi Vahdat[2], Hamed Mirzaiy[3], Azadeh Akhavanallaf[4], Hossein Arabi[4]

[1] Department of Energy Engineering, Sharif University of Technology, Tehran, Iran, Zip code: 8639-11365

[2] Department of Electrical Engineering, Sharif University of Technology, Tehran, Iran, Zip code: 8639-11365

[3] Erfan Radiation Oncology Center (EROC), Tehran, Iran, Zip code: 57176

[4] Division of Nuclear Medicine and Molecular Imaging, Geneva University Hospital, CH-1211 Geneva, Switzerland

**Abstract**

The presence of metal implants within CT imaging causes severe attenuation of the X-ray beam. Due to the incomplete information recorded by CT detectors, artifacts in the form of streaks and dark bands would appear in the resulting CT images. The metal-induced artifacts would firstly affect the quantitative accuracy of CT imaging, and consequently, the radiation treatment planning and dose estimation in radiation therapy. To address this issue, CT scanner vendors have implemented metal artifact reduction (MAR) algorithms to avoid such artifacts and enhance the overall quality of CT images. The orthopedic-MAR (OMAR) and normalized MAR (NMAR) algorithms are the most well-known metal artifact reduction (MAR) algorithms, used worldwide. These algorithms have been implemented on Philips and Siemens scanners, respectively. In this study, we set out to quantitatively and qualitatively evaluate the effectiveness of these two MAR algorithms and their impact on accurate radiation treatment planning and CT-based dosimetry. The quantitative metrics measured on the simulated metal artifact dataset demonstrated superior performance of the OMAR technique over the NMAR one in metal artifact reduction. The analysis of radiation treatment planning using the OMAR and NMAR techniques in the corrected CT images showed that the OMAR technique reduced the toxicity of healthy tissues by 10% compared to the uncorrected CT images.

## 1. Introduction

The high quality of CT images has made it one of the most widely used medical imaging systems in clinical practice [1-5]. The presence of metal implants, such as hip prostheses, metal screws and plates, dental fillings, war fragments, and heart pacemakers can severely attenuate the emitted radiation within CT imaging. Thus, the reconstructed images would bear very poor quality due to the existence of star-shaped artifacts in the form of streaks and dark bands [6]. Metal implants cause severe artifacts in CT images for three main reasons: beam hardening, photon starvation, and scattered photons [7, 8].

The spectrum of the X-ray beams in CT imaging contains relatively low energy photons which are easily absorbed (via photoelectric interactions) when passing through a medium. Therefore, the average energy of the beam increases and the beam hardens while the X-ray beam is passing through a medium; which would significantly alter the overall characteristics of the X-ray spectrum [9, 10]. In the presence of metallic objects, the rate of the absorbed photons radically increases and as a result, fewer photons are detected in the detectors [11]. In this light, the signal-to-noise ratio dramatically drops (due to the increased statistical noise), which leads to image artifacts and poor quality of CT images. As a result of beam hardening (the increased overall energy of the X-ray beam), the scattered photons (Compton scattering) may also noticeably contribute to the signal formation in X-ray detectors. The presence of metal objects causes extraordinary beam hardening (photon starvation), leading to a very poor signal-to-noise ratio and streak-like artifacts in CT images [12]. The other contributory factor is noise which might cause pseudo structures/patterns (noise-induced artifacts) in CT images, particularly when the statistical noise increases due to severe beam attenuation. The detected photons in the presence of metallic objects follow a Poisson distribution, which would cause severe image artifacts because of high signal uncertainty [13-15].

The quantitative accuracy of CT images (CT-derived attenuation coefficients) plays a crucial role in reliable radiation treatment planning [14, 16-19]. Over the past decades, many attempts have been made to resolve/reduce metal artifacts in CT imaging or at least minimize their adverse quantitative impact on CT images mainly through the use of Metal Artifact Reduction (MAR) methods [13, 20]. CT scanner manufactures have also tried to implement/employ the best algorithms to reduce metal-induced artifacts in clinical practice. Philips and Siemens are two major manufacturers of medical imaging systems that have introduced Orthopedic Metal Artifact Reduction (OMAR) and Normalized Metal Artifact Reduction (NMAR) algorithms to their platforms for clinical use [14, 21-23].

In this study, a comparison has been made between two commercial MAR techniques (i.e., OMAR and NMAR) in the context of CT-based radiation planning. To this end, qualitative and quantitative evaluation of the two MAR techniques were performed on a clinical dataset. Standard quantitative image quality metrics such as Normalized Root Mean Square Error (NRMSE), Mean Absolute Deviation (MAD), Mean Relative Error (MRE), and Peak Signal to Noise Ratio (PSNR) were employed to assess the quality of the resulting/corrected CT images. Furthermore, the performance of these two MAR techniques was investigated for CT-based radiation treatment planning (using the Monaco treatment planning software [24]), wherein the accuracy of the recovered CT numbers (after MAR) plays a critical role.

## 2. Materials and methods

### 2.1. MAR techniques

*2.1.A NMAR*

The NMAR technique relies on an image interpolation algorithm to estimate the corrupted/missing regions in CT images, wherein prior knowledge in the form of segmented images into the bone and soft-tissue classes is employed. Via using radon transform, the original images and tissue masks are converted into projection data, then the original sinogram is normalized by dividing the sinogram of the prior image. Finally, the corrupted areas (due to the existing metal artifact) are recovered/predicted using the image interpolation algorithm in the normalized sinogram. The entire framework of the NMAR technique is illustrated in Figure 1.

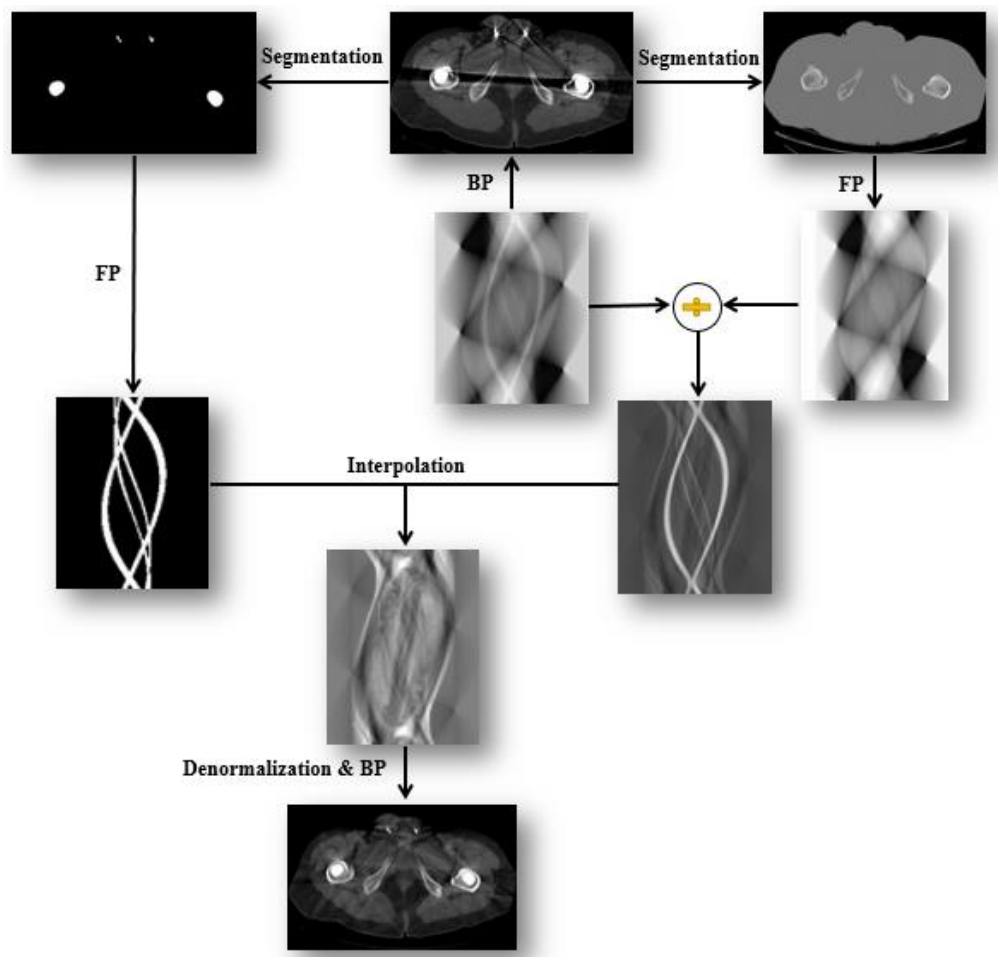

**Figure 1**. Flowchart of the NMAR technique [22].

*2.1.B OMAR*

The OMAR technique is an iterative process wherein the corrected image after each iteration is subtracted from the input (original) image. The new image is regarded as a new input for the next iteration in the algorithm (Figure 2). In the first step, a metal-only image is created via applying an intensity-based threshold on the input image. The rest of the voxels within the body contour are considered as normal tissue with an average value of the entire voxels. This image is then employed to generate the sinogram of only the metal objects in the image (in projection space). If the generated sinogram is empty/zero (no metal voxel), no further processing/iteration is needed; thus, the OMAR technique would have no impact on normal tissues. The input image, as well as the normal tissue image, are forward-projected to create their corresponding sinogram data (projection data). The sinogram data obtained from the normal tissue image is subtracted from the sinogram data obtained from the original image to create an error sinogram. The metal sinogram is then utilized to identify the metal data points in the error sinogram. Thereafter, the resulting error sinogram is back-projected to reconstruct the corrected image. The novelty of this technique lies in the first iteration, wherein all metal data points are removed from the original image. The removed points/pixels are then replaced with the interpolated values using the neighboring normal/unaffected tissue. The output of the first iteration is segmented to create another tissue image to repeat these processes.

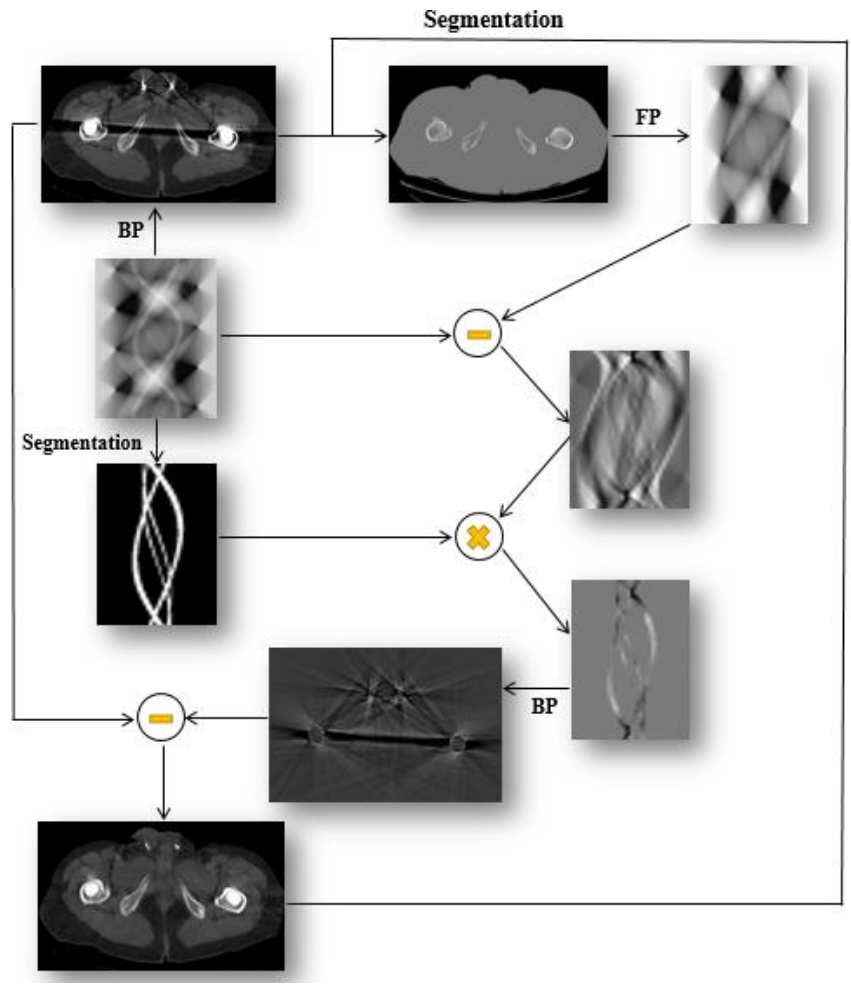

**Figure 2**. Flowchart of the OMAR technique [23].

## 2.2. Metal artifact simulation

The OMAR and NMAR techniques were evaluated using simulated and clinical datasets. Due to the fact that there is no ground truth for the clinical dataset, metal artifact simulation was also considered to evaluate MAR techniques performance [25]. To this end, a metal artifact simulation framework, described in [26], was employed to mimic those metal artifacts in CT images without any metal implants. First, the CT values were converted to the attenuation coefficients. Then, the attenuation coefficients corresponding to the metallic objects were inserted in the CT images, and next, the projection data was generated using the radon transform. The simulated projection data was then reconstructed using the filter back-projection algorithm to generate the CT images with metal artifacts. To simulate metal artifacts in CT images, we employed the algorithm/source code developed by Sakamoto et al. which is described in [26]. A representative example of the inserted metal implant and the simulated metal artifact is depicted in Figure 3.

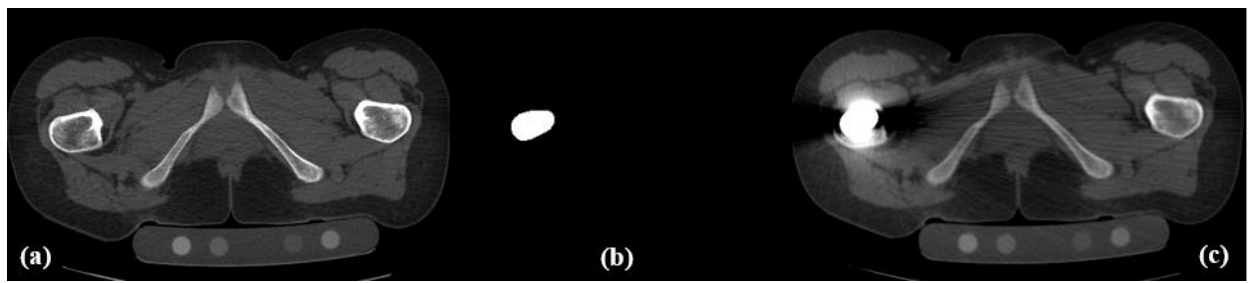

**Figure** 3. (a) Metal-free CT image (reference), (b), inserted metal implant, and (c) simulated metal artifact.

**2. 3. Evaluation**

*2. 3. 1. Quantitative metrics*

Quantitative analysis of these two MAR techniques was performed through the calculation of the normalized squared error (NRMSE), mean absolute deviation (MAD), mean relative error (MRE), and peak signal to noise ratio (PSNR) using Eqs. (1)-(4), respectively. These parameters were calculated between the simulated CT images with metal artifacts and the corresponding reference artifact-free images [25, 27, 28].

$$NRMSE = \sqrt{\frac{\sum_{J=1}^{N}(I_J^{corr} - I_J^{ref})^2}{\sum_{J=1}^{N}(I_J^{ref} - I^{ref})^2}} \quad (1)$$

$$MAD = \frac{1}{N}\sum_{J=1}^{N}\left|I_J^{corr} - I_J^{ref}\right| \quad (2)$$

$$MRE = \left| \frac{I^{corr} - I^{ref}}{I^{ref}} \right| \qquad (3)$$

$$PSNR = 10 \times log \left( \frac{I_{max}^2}{\left\| I_J^{corr} - I_J^{ref} \right\|^2} \right) \qquad (4)$$

where, $I_J^{corr}$ and $I_J^{ref}$ are the CT numbers of the corrected and reference images (artifact-free), respectively. $I^{corr}$ and $I^{ref}$ denote the average CT number of pixels in the corrected and reference images, respectively. $N$ indicates the total number of pixels and $I_{max}$ is the maximum pixel value.

**2. 3. Dosimetry**

Dose calculation requires attenuation coefficient maps of the body, which are normally derived from CT images. These coefficient maps enable quantitative calculation of the depth of radiation penetration into the biological tissues, such as the water-equivalent thickness (WET) of the material according to Eq. (5).

$$WET = d_m \rho \qquad (5)$$

where, $d_m$ is the tissue thickness and $\rho$ is the electron density relative to water (in photon radiation therapy), derived from CT images. The electron density maps obtained from the CT images before and after MAR were used to simulate the dose distribution. Calculation of dose distribution was carried out using different computational methods, such as the Analytical Anisotropic Algorithm (AAA) provided by the Eclipse Treatment Planning System (TPS) [29].

Under normal circumstances, small errors in the estimation of the $\rho$ from CT images would lead to dose calculation errors of less than 1% [30]. In the presence of metal implants, however, depending on the severity of metal artifacts, the estimated dose in treatment planning may increase up to 10% [31].

**3. Results and discussion**

*3.1. Quantitative metrics*

In this study, OMAR and NMAR algorithms were applied to head and femur CT images, wherein the dental fillings and metal implants cause severe metal artifacts. Figures 4 and 5 show representative slices of the corrupted CT images together with the corrected images by the OMAR and NMAR techniques. Figure 4 shows the metal artifacts resulted from femur and hip implants, wherein both the OMAR and NMAR techniques greatly reduced the metal artifacts without introducing secondary artifacts to the images. In general, the OMAR approach resulted in slightly sharper recovery of the underlying structures; however, in comparison with the NMAR approach, it failed to completely remove the metal-induced artifact in Figure 4. slice 3.

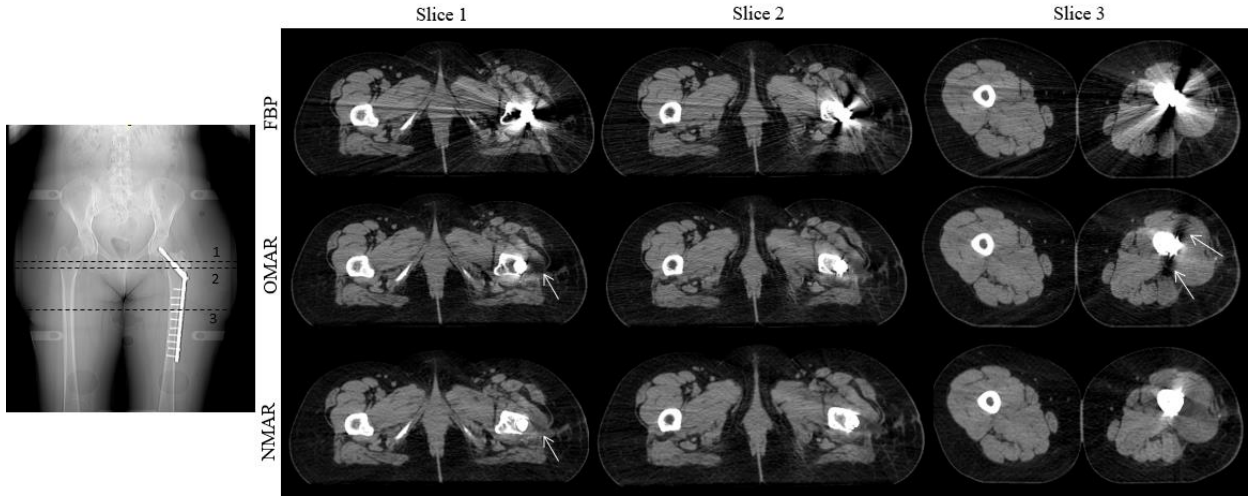

**Figure 4.** Representative slices of the corrupted CT image with the femur and hip implants together with corrected images with the OMAR and NMAR techniques.

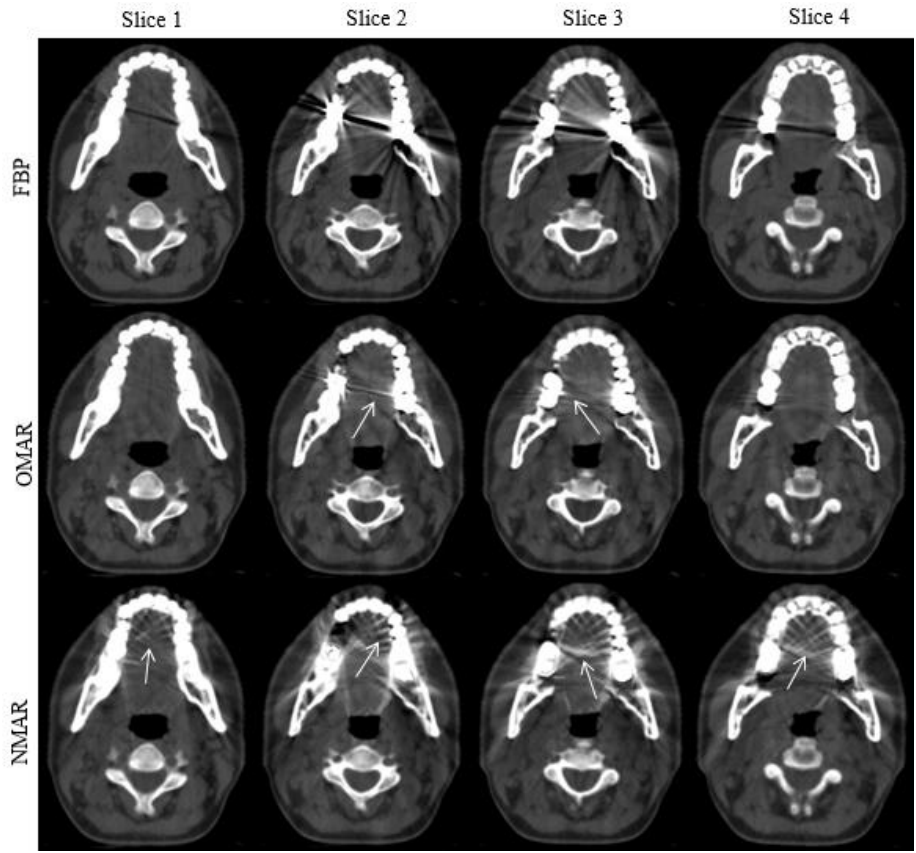

**Figure 5.** Representative slices of the corrupted CT image with dental fillings together with corrected images with the OMAR and NMAR techniques.

Figure 5 depicts the strong streak artifacts due to dental fillings, wherein the NMAR approach introduced secondary artifacts within the affected areas. The NMAR algorithm relies on the segmentation of the input image into few major tissue classes as prior knowledge [22]; thus, the existence of severe metal artifacts would lead to the mis-segmentation of the input image. The mis-segmentation of the input image would provide the NMAR algorithm with inaccurate prior knowledge which would result in secondary artifacts in the corrected images. In this regard, the OMAR approach outperformed the NMAR approach, wherein no and/or less secondary artifacts were observed. Similar to Figure 4, the dark streaks in slices 2 and 3 were not completely removed/corrected by the OMAR algorithm. In general, the OMAR approach exhibited superior performance in MAR for the head CT images.

To evaluate the accuracy of the estimated CT values after MAR, the line profiles of the corrected images by the two approaches were compared against the reference image. To this end, the simulated metal artifact data were used, wherein ground truth images were available. Figure 6. (a) illustrates the line profiles drawn on the original (corrupted) image, the reference image, and the corrected images by the two approaches. Figure 6. (b) illustrates the absolute error of the line profiles for the OMAR, NMAR, and the original image (corrupted image) versus the line profile of the reference image. The absolute error is calculated using Eq. (6).

$$AE(u) = |CT_{Corrected}(u) - CT_{Reference}(u)| \tag{6}$$

The OMAR algorithm recovered the CT values slightly better than the NMAR algorithm showing closer agreement with the reference images. In Figure 6.(b), the error due to the metal artifact is clearly seen in the line profile of the original (corrupted) image.

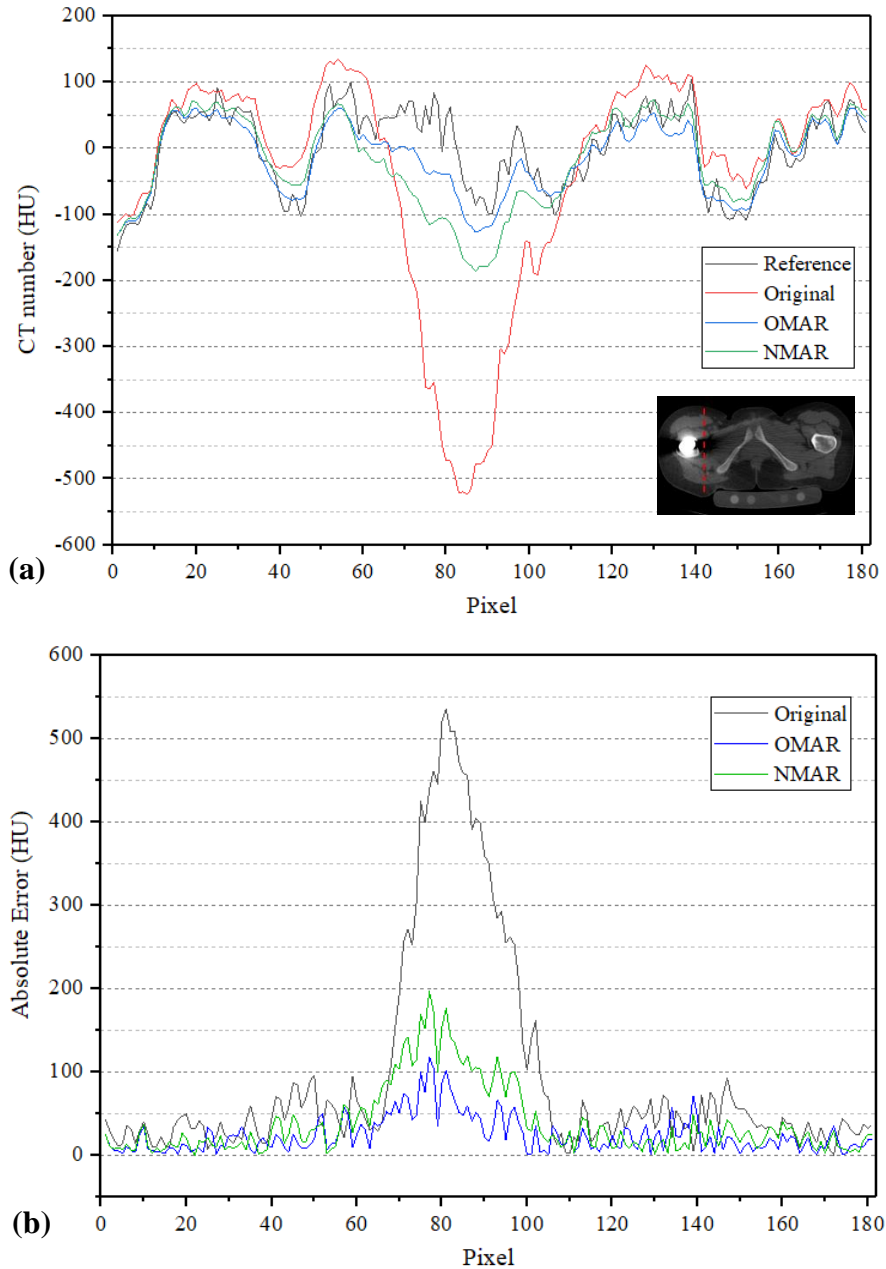

**Figure 6.** Comparison of the line profiles drawn on the original (corrupted) image, reference image, and corrected images by the NMAR and OMAR approaches (a). Absolute difference of the profiles versus the reference profiles (b).

Table 1 summarizes the NRMSE, MRE, MAD, and PSNR parameters calculated for the original (corrupted) images together with the corrected ones with the OMAR and NMAR algorithms. For the corrected images by the OMAR algorithm, the NRMSE, MRE, MAD, and PSNR of 0.21, 1.91, 47.58, and 30.53 were observed, respectively. Considering the PSNR metric, 8 and 6 dB improvement was observed when using the OMAR and NMAR algorithms, respectively. Regarding the quantitative metrics, overall, the OMAR algorithm exhibited superior performance to the NMAR algorithm leading to smaller NRMSE and MRE.

**Table 1.** Quantitative metrics calculated for the OMAR and NMAR algorithms on the simulation dataset.

| Method | NRMSE | MRE | MAD | PSNR |
|---|---|---|---|---|
| **FBP** | 0.88 | 5.56 | 82.22 | 22.21 |
| **OMAR** | 0.21 | 1.91 | 47.58 | 30.53 |
| **NMAR** | 0.24 | 2.49 | 55.44 | 28.75 |

*3.2. Dosimetry*

The quantitative accuracy of CT images has a great influence on dose calculation and isodose curves estimation in radiation therapy [32]. Dose distribution in conformal plans with two posterior and anterior oblique beams with 6 MeV beam energy was performed on the original (corrupted) CT images as well as the corrected ones by both algorithms. Figure 7 compares the dose distribution calculated on the original (corrupted) and corrected CT images by the two algorithms for a patient with dental fillings.

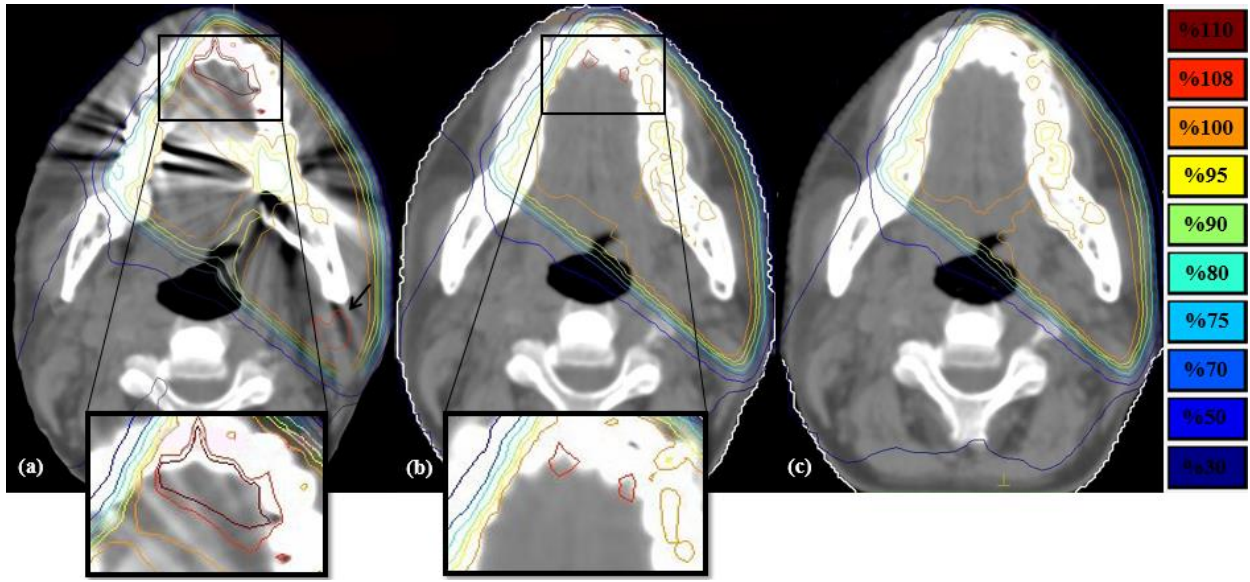

**Figure 7.** Isodose curves for two oblique beams of 6 MeV and 2 Gy prescribed dose. Dose distribution calculated on the (a) original (corrupted) CT images, (b) CT image corrected by the NMAR algorithm, and (c) CT images corrected by OMAR algorithm.

Due to the severe metal artifacts, it is very challenging to correctly segment/detect tissues near the filled teeth in the original image (corrupted). On the other hand, due to the errors in CT numbers, it is not possible to correctly derive the electron density for the accurate estimation of the dose distribution in the corrupted image. In this light, areas with an absorbed dose of more than 110% were observed in the corrupted images, which are called hot spots [29]. The hot spots in isodose curves not only cause unnecessary doses to healthy/normal tissues around the target area but also disrupt the dose uniformity within the tumor or target area.

The application of the NMAR algorithm improved the quantitative accuracy of the CT images and consequently reduced their dose estimation errors. However, the dose estimation error resulted from the remaining artifacts close to severe metal artifacts in the CT images, led to noticeable WET estimation errors. Much fewer hot spots were observed after the correction by the NMAR algorithm; however, the presence of these hot spots would still lead to errors in radiotherapy and the increase of unnecessary doses to healthy tissues. The red curves in Figure 7 represent the depth doses of more than 110%, which can be seen in the isodose curves for the original (corrupted) CT image and the corrected CT image by the NMAR algorithm. Moreover, the relatively high accuracy of the OMAR algorithm in recovering the correct CT values led to an accurate dose estimation. As shown in Figure 7.(c), the 110% isodose curve was completely eliminated and the dose for sensitive organs and healthy tissues was reduced by nearly 10%.

A review of recent similar studies shows that previous researchers have reached similar conclusions to the present study, wherein the promising ability of the OMAR algorithm to recover correct CT values in corrupted CT images by metal artifacts was reported. The effectiveness of the OMAR algorithm and its positive effects on radiation treatment planning was demonstrated via experimental and clinical studies by Li et al. [33]. Li et al. showed that the OMAR algorithm is able to recover the underlying anatomical structures in corrupted images by metal artifacts. In Bolstad et al. [34] research, the performance of four commercial MAR algorithms on the images corrupted by metal artifacts due to different metal implants was rated by radiologists, wherein the OMAR algorithm was ranked higher than other algorithms in almost all cases.

## 4. Conclusion

In this study, we compared the performance of two commercially available MAR algorithms using clinical hip and head CT images as well as a simulated metal artifact dataset. The performance of the MAR algorithms was also assessed in dose estimation and isodose calculation for radiation treatment planning. Overall, the quantitative evaluation demonstrated superior performance of the OMAR algorithm versus the NMAR algorithm in metal artifact reduction in clinical as well as simulation datasets. Investigation of dose distribution on the corrected CT images by the OMAR and NMAR algorithms showed superior performance of the OMAR algorithm to reduce the unnecessary dose to healthy tissues by 10%.